\documentclass[aps,prd,twocolumn,showpacs,preprintnumbers,amsmath,amssymb,
epsf,superscriptaddress,groupedaddress, tightenlines]{revtex4}
\usepackage{epsfig,graphicx}
\usepackage{graphicx}
\usepackage{amsmath}
\usepackage{bm}
\newcommand{\nn}{\nonumber}
\newcommand{\vslash}{v\hspace*{-5.5pt}\slash}

\newcommand{\Dslash}{D\hspace*{-5.5pt}\slash}
\newcommand{\varepsslash}{\varepsilon\hspace*{-5.5pt}\slash}
\newcommand{\epsslash}{\varepsilon\hspace*{-5.5pt}\slash}
\newcommand{\peslash}{p_e\hspace*{-8.5pt}\slash}
\newcommand{\qslash}{q\hspace*{-5pt}\slash}
\newcommand{\pslash}{p\hspace*{-5.5pt}\slash}

\def\nslash{n\!\!\!\slash}
\def\bnslash{\bar n\!\!\!\slash}
\def\bn{\bar n}

\begin{document}
\preprint{MIT-CTP 3672}
\preprint{CALTECH-MAP-308}

\title{%\mbox{}\\[4pt]
Factorization in exclusive semileptonic radiative 
$B$ decays}



\author{Vincenzo Cirigliano}
\affiliation{California Institute of Technology, Pasadena, CA 91125}

\author{Dan Pirjol}
\affiliation{Center for Theoretical Physics, Massachusetts Institute for
  Technology, Cambridge, MA 02139}
%\thanks{}


%\date{\today}

\begin{abstract}
We derive a new factorization relation for the semileptonic radiative 
decay $B\to \pi \ell \nu \gamma$ in the kinematical region of a slow
pion $|\vec p_\pi\, | \sim \Lambda$ and an energetic photon 
$E_\gamma \gg \Lambda$, working at leading order in $\Lambda/m_b$.
In the limit of a soft pion, the nonperturbative matrix element appearing in 
this relation can be computed using chiral perturbation theory. We present 
a phenomenological study of this decay, which may be important for a precise 
determination of the exclusive nonradiative decay.
\end{abstract}

%12.39.Fe Chiral Lagrangians
%13.60.-r photon-hadron interaction

%\keywords{}

\maketitle

%\tighten
%\section{Introduction}
%\subsection{introduction}
%\subsubsection{}

\section{Introduction}

Experiments at the B factories are reaching an accuracy level that
requires a good understanding of radiative effects in weak B
decays. In particular, real photon emission distorts the spectra of
other decay products, therefore directly affecting the distribution of
kinematic quantities used to select and count signal
events~\cite{RadCorr,CGatti}. As a consequence, real photon emission
indirectly affects branching fraction measurements, leading to
systematic shifts of the order of a few percent.  While radiative
effects are well studied in semileptonic and nonleptonic kaon decays,
where chiral perturbation theory is applicable over the entire
kinematical range (see Refs.~\cite{kl31,kl32,D'Ambrosio:1994ae}), the
situation is less satisfactory in B decays. Here the large energy
release generally prohibits the application of chiral perturbation
theory methods over large parts of the phase space, and one has to
resort to different theoretical tools.

We study here one of the simplest radiative processes, the semileptonic
radiative B decays $B\to \pi \ell \nu \gamma$. The corresponding
nonradiative process, the semileptonic decay $B\to \pi\ell \nu$, is
well studied in connection with the determination of the CKM matrix
element $|V_{ub}|$. The measured branching fractions of this mode 
%from the B factories 
are \cite{exp}
\begin{eqnarray}
&& {\cal B}(B^0\to \pi^- \ell^+\nu) =\\ 
&& \left\{
\begin{array}{cc}
(1.33\pm 0.18 \pm 0.13) \times 10^{-4} & \mbox{(CLEO)} \\
(1.38\pm 0.10 \pm 0.18) \times 10^{-4} & \mbox{(BABAR)} \\
(1.48\pm 0.20 \pm 0.16) \times 10^{-4} & \mbox{(BELLE)} \\
\end{array}
\right.
\nonumber
\end{eqnarray}
which contribute to the world average $B(B^0\to \pi^- \ell^+\nu) = 
(1.35\pm 0.08 \pm 0.08) \times 10^{-4}$
\cite{hfag}.

The hadronic dynamics of the radiative semileptonic decays is
complicated by the presence of many energy scales. However, in certain
kinematical regions a model independent treatment is possible.  In the
limit of a soft photon, Low's theorem \cite{Low} fixes the terms
of order $O(q^{-1})$ and $O(q^0)$ in the expansion of the decay
amplitude in powers of the photon momentum $q$. These contributions
have been computed in \cite{fearing} in kaon semileptonic radiative
decays $K_{\ell 3}$. Higher order terms in $q$ have been more recently
obtained using chiral perturbation theory \cite{kl31,kl32}.  A similar
approach is possible in B decays with the help of the heavy hadron
chiral perturbation theory \cite{wise,BuDo,TM,review}, as long as the
pion and photon are sufficiently soft.  This is the case in a limited
corner of the phase space, corresponding to a large dilepton invariant
mass $W^2=(p_{e}+p_\nu)^2 \sim m_b^2$.  

The radiative semileptonic decay $B\to \gamma \pi \ell \bar\nu_\ell$
at low $W^2$
produces in general either an energetic pion or an energetic photon (see Fig.~1).
In this paper we use soft-collinear effective theory 
(SCET)~\cite{scet0,bfps,scet2,scet3}-\cite{scet3.5,scet4} methods to 
prove a factorization relation for the decay amplitude {\em in the kinematical
region with an energetic photon and a soft pion}. This can be written
schematically as
\begin{equation}\label{1}
A(B\to \pi \ell \nu \gamma) = H \otimes J \otimes S(B\to \pi) +
O(\Lambda/Q)
\end{equation}
with $\Lambda \sim \Lambda_{QCD}$ and 
$Q \sim \{E_\gamma , m_b \}$. Here $H$ and $J$ are hard and jet
functions, calculable in perturbation theory. $S(B\to \pi)$ is related
to the matrix element of a nonlocal soft operator which describes the
$B\to \pi$ transition. 
This factorization relation generalizes to any process $B\to
\gamma M\ell \nu$ ($M = \rho, \omega, \cdots$ a light soft meson) 
with the appropriate soft function $S(B\to M)$ replacing $S(B\to
\pi)$.
In this work we use heavy hadron chiral perturbation theory to compute
the soft matrix element $S(B\to \pi)$, and express it in terms of one
of the B meson light-cone wave functions
$\phi_+^B(k_+)$~\cite{GPchiral}.  The convolution of this function
with the jet function $J$ in Eq.~(\ref{1}) is similar to the one
appearing in a factorization relation for the radiative leptonic decay
$B\to \gamma \ell\nu$ \cite{radlept0,radlept1,radlept2,radlept3}.

The paper is organized as follows: after fixing the notation in
Section~\ref{sect:definitions}, we describe the main results of the
paper (factorization formula for hadronic form factors) in
Section~\ref{sect:results}.  We give a proof of the factorization
formula in Section~\ref{sect:proof}.  In Section~\ref{sect:pheno} we
explore the phenomenological consequences of our results, computing the
radiative branching fractions with appropriate cuts on the photon and
pion energy, as well as various differential decay distributions.  The
factorization results are compared with the ones obtained by
extrapolating Low's amplitude to the region of an energetic photon.
We collect our conclusions in Section~\ref{sect:conclusions}.  An
appendix contains the expression of the radiative amplitude in Low's
limit of a soft photon.

\section{Amplitude definitions}
\label{sect:definitions}

We consider in this paper the radiative semileptonic B decays
\begin{eqnarray}
\bar{B}(p) \to \pi(p') \,  
e^-(p_e) \, \bar{\nu}(p_\nu) \, \gamma(q,\varepsilon)  \ , 
\end{eqnarray}
mediated by the combined semileptonic weak Hamiltonian 
\begin{eqnarray}
{\cal H}_W = \frac{4G_F}{\sqrt2} V_{ub} [\bar u \gamma_\mu P_L b]
[\bar e \gamma^\mu P_L \nu]
\end{eqnarray}
and the electromagnetic coupling of quarks and charged leptons. 

The full amplitude for $\bar{B}(p) \to \pi(p') e^-(p_e)
\bar{\nu}(p_\nu) \gamma(q,\varepsilon)$ contains two terms, describing
the photon coupling to the hadronic system and to the charged lepton,
respectively
\begin{eqnarray}\label{full}
A & = & \frac{e G_F}{\sqrt{2}} V_{u b} \varepsilon^\mu(q)^{*}  
\Big[  \left( T_{\mu \nu}^{V} -  T_{\mu \nu}^{A} \right) \, 
\bar{u} (p_e) \gamma^\nu (1 - \gamma_5) v(p_\nu)    
\nonumber \\
&+&  \frac{F^\nu (p,p')}{2 \, q \cdot p_e} \, \bar{u} (p_e)  \gamma_\mu
(m_e + \peslash + \qslash) \gamma_\nu (1 - \gamma_5) v (p_\nu) 
\Big] \ .  \nonumber \\
& & 
\end{eqnarray}
We denoted here
\begin{equation}\label{FFdef}
F_\nu (p,p') = \langle \pi(p')| \bar{u} \gamma_\nu b |\bar{B}(p) \rangle  \ , 
\end{equation}
and the hadronic tensors $T_{\mu \nu}^{V,A}$ are defined as
\begin{equation}
T^{J}_{\mu\nu} = i
\int d^4 x e^{iq\cdot x}  \langle
\pi(p') |\mbox{T} \left\{ j_\mu^{\rm em}(x)\,, J_\nu(0) \right\} 
|\bar B(p)\rangle
\label{TmunuJ}
\end{equation}
with $j_\mu^{\rm em}=\sum_q e_q \bar q\gamma_\mu q$ the quarks' electromagnetic
current and:
%The weak currents relevant for semileptonic decays are 
%$J_\nu = V_\nu - A_\nu$.
%The corresponding hadronic tensors are denoted as
\begin{eqnarray}
&& T^V_{\mu\nu}\,: \qquad J_\nu = V_\nu = \bar u \gamma_\nu b\\ &&
T^A_{\mu\nu}\,: \qquad J_\nu = A_\nu = \bar u \gamma_\nu\gamma_5 b\, .
\end{eqnarray}
The hadronic tensors satisfy the Ward identities
\begin{eqnarray}
&& q^\mu T^{V}_{\mu\nu} = 
%
% - \langle \pi(p')|V_\nu |B(p)\rangle \equiv 
%
- F_\nu(p,p')\\
&& q^\mu T^{A}_{\mu\nu} = 0 \ . 
%&& q_\mu T_{\mu\nu} = ?
\end{eqnarray}

The most general form for the vector amplitude compatible with the
Ward identity can be written in terms of 4 invariants $V_{1-4}$. After
multiplication with the photon polarization, this is written as
($W = p_e + p_\nu$)
\begin{eqnarray}
&& \varepsilon^{*\mu} T^{V}_{\mu\nu} = V_1 (\varepsilon_\nu^* - \frac{\varepsilon^*\cdot W}
{q\cdot W}q_\nu )\\
&& + (p'\cdot \varepsilon^* - \frac{(p'\cdot q)(\varepsilon^*\cdot W)}{q\cdot W}) 
(V_2 q_\nu + V_3 W_\nu + V_4 p'_\nu)\nn\\
&& \qquad - \frac{\varepsilon^*\cdot W}{q\cdot W}F_\nu(p,p')\nn\,.
\end{eqnarray}

The axial amplitude $T^{A}_{\mu\nu}$ can be written in terms of  
another four invariants $A_{1-4}$. They can be chosen as 
\begin{eqnarray}
 \varepsilon^{*\mu} T^{A}_{\mu\nu} &=& 
i\epsilon_{\mu\nu\rho\sigma}\varepsilon^{*\mu} 
(A_1 p'_\rho q_\sigma + A_2 q_\rho W_\sigma) \\
&+& 
i\epsilon_{\mu\lambda\rho\sigma} \varepsilon^{*\mu} p'_\lambda q_\rho W_\sigma
(A_3 W_\nu + A_4 p'_\nu) \nn
\end{eqnarray}

The invariant form factors $V_{1-4}$ and $A_{1-4}$ depend on 3 independent kinematical
variables, which can be chosen as $W^2, E_\pi, E_\gamma$ with $E_\pi$ and $E_\gamma$
defined as the pion and photon energy in the rest frame of the B meson. For given $W^2$,
the possible final states in the decay can be represented as points in the Dalitz plot 
for $(E_\gamma, E_\pi)$. We show in Fig.~1 the Dalitz plot for $W^2 = 4$ GeV$^2$.
The dilepton invariant mass $W^2$ can take values $W^2=[0,(m_B-m_\pi)^2] = [0,26.5]$ GeV$^2$.

\begin{figure}[b!]
\begin{tabular}{cc}
{\includegraphics[height=5cm]{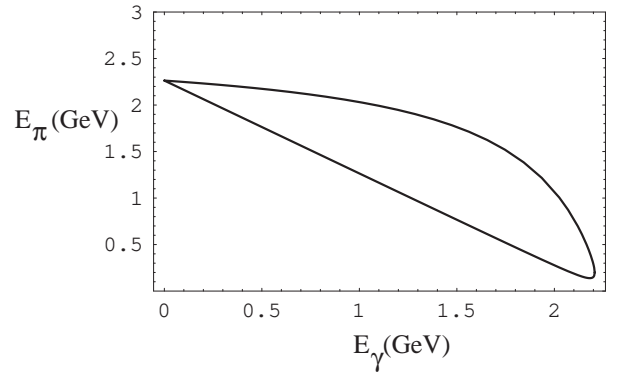}}
\end{tabular}
\caption{\label{fig1} 
The Dalitz plot for $B\to \gamma\pi \ell\bar\nu$ at $W^2=4$ GeV$^2$.
Our computation of the form factors is applicable in the lower right corner of
the Dalitz plot, with an energetic photon and a soft pion.
}
\end{figure}

The form factors in $B$ (quark content $\bar b q$) and $\bar B$ ($b\bar q$ quark content)
are related by charge conjugation. The transformations of the currents under charge
conjugation are
\begin{eqnarray}
{\cal C} \bar q \gamma_\nu b {\cal C}^\dagger = - \bar b \gamma_\nu q \,,\quad
{\cal C} \bar q \gamma_\nu\gamma_5 b {\cal C}^\dagger = \bar b \gamma_\nu\gamma_5 q
\end{eqnarray}
We adopt everywhere in this paper the following phase conventions for the meson states
\begin{eqnarray}
&& (\pi^+\,,\pi^0\,,\pi^-) = (u\bar d, \frac{1}{\sqrt2}(u\bar u - d \bar d),  d\bar u)\\
&& (B^-, \bar{B}^0 ) = (b\bar u, b \bar d)\,,\quad
( B^0, B^+ ) = (\bar b d, \bar b u)\,.
\end{eqnarray}
With this convention the states transform as $ {\cal C} \pi^+ = \pi^-\,, 
{\cal C} \bar B^0 =  B^0\,,{\cal C} B^- =  B^+ $. In addition, the $\pi^0$ is
C-even and the photon is C-odd.
The semileptonic radiative $B\to \pi$ form factors transform as
follows under charge conjugation, independent of the flavor of 
the spectator quark
\begin{eqnarray}\label{radslC}
&& V_{1-4}(\bar B \to \pi \gamma \ell^- \bar\nu_\ell) = 
 V_{1-4}(B \to \pi \gamma \ell^+ \nu_\ell) \\
&& A_{1-4}(\bar B \to \pi \gamma \ell^- \bar\nu_\ell) = 
- A_{1-4}(B \to \pi \gamma \ell^+ \nu_\ell)
\end{eqnarray}
The vector form factor $F_\nu(p,p')$ is C-odd. 
Finally, the form factors depend also on the flavor of the light spectator 
in the B meson. For simplicity of notation, we chose to omit this dependence, 
but it will be specified explicitly whenever required.

\section{Results}
\label{sect:results} 

We describe here the main results of this paper.  In the kinematical
region $E_\gamma \gg \Lambda$ and $E_\pi \sim \Lambda$ (hard photon
and soft pion), the form factors for $B\to \pi \ell\nu \gamma$ are
given by a factorization theorem. At leading order in $\Lambda/Q$ with
$Q \sim \{m_b, E_\gamma\}$, these factorization relations are the same
for both $B^-, \bar B^0$, up to an isospin factor.  
The predictions for the vector form factors 
in $\bar B^0\to \pi^+ \ell^-\bar \nu \gamma$ are
\begin{eqnarray}\label{V14}
&& V_1 = 2e_u C_1^{(v)}(E_\gamma) \int dk_+ J(k_+) S_1(k_+, t^2, \zeta) \\
&& V_2 = \frac{e_u}{E_\gamma} \Big( 2C_1^{(v)}(E_\gamma) + 
C_2^{(v)}(E_\gamma) + 2C_3^{(v)}(E_\gamma) \Big) \nn \\
&& \qquad \qquad \times  \int dk_+ J(k_+) S_2(k_+, t^2, \zeta)  \\
&& V_3 = \frac{2e_u}{n\cdot W} C_2^{(v)}(E_\gamma) \int dk_+ J(k_+) S_2(k_+, t^2, \zeta) \\
&&  V_4/V_1 \sim O(\frac{1}{\Lambda Q})
\end{eqnarray}
and the corresponding relations for the axial form factors are
\begin{eqnarray}\label{A14}
&& A_1 = \frac{e_u}{E_\gamma} \Big( 
2C_1^{(v)}(E_\gamma) + C_2^{(v)}(E_\gamma) + 
2C_3^{(v)}(E_\gamma) \Big) \nn \\
&& \qquad \qquad \times \int dk_+ J(k_+) S_2(k_+, t^2, \zeta)
  \\
&& A_2 = \frac{2e_u}{q\cdot W} C_1^{(v)}(E_\gamma) \int dk_+ J(k_+) S_1(k_+, t^2, \zeta)
 \\
&& A_3 = \frac{2e_u}{q\cdot W n\cdot W} C_2^{(v)}(E_\gamma) \\
&& \qquad \qquad \times
\int dk_+ J(k_+) S_2(k_+, t^2, \zeta)  \nn \\
\label{A4}
&& A_{4}/A_2 \sim O(\frac{1}{\Lambda Q}) \,.
\end{eqnarray}
After multiplication by the appropriate tensor structures, $V_4$ and
$A_4$ give subleading contributions in $\Lambda/Q$ to the hadronic
tensor.  The corresponding form factors for $B^-\to \gamma \pi^0
\ell^-\bar\nu$ are multiplied by a factor $1/\sqrt2$.  In these
formulas $n_\mu = q_\mu/E_\gamma$ is a light-cone unit vector along
the photon momentum.

The hard functions $C_{1-3}^{(v)}(E_\gamma)$ are Wilson coefficients
in the SCET, and are calculable in perturbation theory in an expansion
in $\alpha_s(Q)$ as $C_1^{(v)} = 1 + O(\alpha_s(m_b))\,, C_{2,3}^{(v)}
= O(\alpha_s(m_b))$.  Their one-loop expressions can be found in
Ref.~\cite{bfps}. The jet function $J(k_+)$ is calculable in an
expansion of $\alpha_s(\sqrt{Q\Lambda})$ and its one-loop expression
is given below in Eq.~(\ref{Jdef}). Finally, $S_{1,2}(k_+, t^2,\zeta)$
are nonperturbative soft matrix elements, defined below in
Eqs.~(\ref{H1def}), (\ref{H2def}).

The results (\ref{V14})-(\ref{A4}) show that, at leading order in
$\Lambda/Q$, the eight form factors parameterizing $B\to \gamma\pi
e \nu$ are given by only two nonperturbative parameters and thus
satisfy symmetry relations.

The soft functions $S_{1,2}$ depend in general on 
$k_+$, $t^2 = (p - p')^2$, $\zeta = n \cdot p' /n \cdot p$, 
and  are calculable at leading order in chiral perturbation theory. 
In the soft pion limit
they can be expressed in terms of one of the $B$ meson
light-cone wave function and are given by
\begin{eqnarray}
 S_1(k_+, t^2, \zeta) &=& - \frac{f_B m_B}{4 f_\pi}  \phi_+^B(k_+)
\left( 1 + g \frac{e_3\cdot  p'} {v\cdot p' + \Delta} \right)
\ \ \ \ 
\label{S1res} \\
 S_2(k_+, t^2, \zeta) &=& \frac{ g f_B m_B}{4 f_\pi} \phi_+^B(k_+) 
\frac{1}{v\cdot p' + \Delta} 
\label{S2res}
\end{eqnarray}
where $v=p/m_B$, $\Delta = m_{B^*} - m_B = 50$ MeV, $e_3= n - v$,  
and $g$ is the axial effective coupling defined in
Eq.~(\ref{Lag}).
To this order, 
the dependence on 
nonperturbative dynamics of the B meson enters only through one
quantity, the convolution of the light-cone B wave function $\phi_+^B(k_+)$
with the jet function $J(k_+)$:  
\begin{eqnarray}\label{Idef}
V_i \, , \, A_i \  \propto  \int dk_+ J(k_+) \phi_B^+(k_+,\mu) \ .
\end{eqnarray}
A similar convolution appears in the factorization relation for the
radiative leptonic $B$ decay $B\to \gamma e\bar \nu$, the only
difference being that in this case the argument of the jet function is
time-like ($+ 2E_\gamma k_+$) as opposed to spacelike in the radiative
leptonic case. As a consequence the form factors $V_i$ and $A_i$ are in
general complex, receiving strong phases of $\sim
\alpha_s(\sqrt{2E_\gamma \Lambda})$ via the loop corrections to the
jet function $J$. 

\section{Factorization proof}
\label{sect:proof}

The proof of the relations Eqs.~(\ref{V14})-(\ref{A4}) makes use of
the soft collinear effective theory (SCET)
\cite{scet0,bfps,scet2,scet3,scet3.5,scet4}.  
The basic idea will be
to match the correlator of currents in Eq.~(\ref{TmunuJ}) onto soft
SCET operators, and then in a second step to construct the chiral
representation of the resulting soft operators. The $B \to \pi$ matrix
element of the correlator in Eq.~(\ref{TmunuJ}) will be computed in
the low energy chiral perturbation theory.

We start by setting up the kinematics of the process. We introduce two light cone unit vectors
$n,\bar n$ with $n^2 = \bar n^2 = 0\,, n\cdot \bar n = 2$. The dilepton momentum $W$ is chosen along 
$\bar n$
with light cone coordinates $W = (n\cdot W, \bar n \cdot W, W_\perp)$ 
(such that $n\cdot W \gg \bar n\cdot W, W_\perp$).
The photon is emitted along $n_\mu$, such that its light-cone coordinates are
$q = (0, \bar n\cdot q,0)$, and the pion momentum is soft, with components
$p_\pi \sim \Lambda$.

We are interested in the correlator of the weak semileptonic current with the 
electromagnetic current 
\begin{eqnarray}\label{corr}
T_{\mu\nu}(q) =
i\int d^4 x e^{iq.x} T \{(\bar q\gamma_\nu P_L b)(0)\,, j_\mu^{\rm e.m.}(x) \}
\end{eqnarray}
The matrix elements of this correlator define the hadronic tensors in
the decays of interest Eqs.~(\ref{TmunuJ}).  The
flavor dependence of the form factors follows from the transformation
properties of this correlator under flavor SU(3). This is given by
$\overline{\mathbf 3} \otimes {\mathbf 8} = \overline{\mathbf 3} + 
{\mathbf 6} + \overline{\mathbf 15}$, so that 
the $\bar{B} \to \pi$ transitions are in general described by three
unknown reduced matrix elements.  However, at leading order in
$\Lambda/E_\gamma$, only one amplitude contributes.   
To see this, consider the two types of contributions to the correlator
% There are two types of contributions to the correlator
$T_{\mu\nu}(q)$, corresponding to the photon attaching either to the
current quarks ($u$ and $b$), or the spectator quark. At leading order
in $\Lambda/E_\gamma$, only the first amplitude contributes, while the
spectator effect is power suppressed (see below).
The dominant amplitudes transform as a
$\overline{\mathbf 3}$ under flavor SU(3), so they are independent of
the flavor of the spectator quark.

The semileptonic $b\to u$ current is first matched onto SCET$_I$, an
effective theory containing ultrasoft quarks and gluons with momenta
$p_s \sim \Lambda$, and collinear modes with virtuality $p_c^2 \sim
Q\Lambda$ moving along $q_\mu \sim n_\mu$ \cite{bfps}.  Neglecting
light quark masses, the $V-A$ current is matched at leading order in
$\lambda \equiv \Lambda/Q$ only onto structures containing one left-handed light quark
field \cite{bfps}
\begin{eqnarray}\label{Jweak}
&& \bar q \gamma_\mu P_L b = 
C_1^{(v)}(\omega) \bar q_{n,\omega} \gamma_\mu P_L b_v\\
&& + 
(C_2^{(v)}(\omega) v_\mu + C_3^{(v)}(\omega) \frac{n_\mu}{n\cdot v}) \bar q_{n,\omega}  P_R b_v 
+ O(\lambda)\nn
\end{eqnarray}
where $C_i^{(v)}(\omega)$ are Wilson coefficients calculable in
perturbation theory as an expansion in $\alpha_s(Q)$.

We match first the correlator Eq.~(\ref{corr}) onto SCET$_{\rm I}$ operators. 
The weak current is matched as shown above in Eq.~(\ref{Jweak}). The light quark 
e.m. current 
coupling $\bar q {\cal A}_n \xi_n$
of an energetic ($n-$collinear) photon to a ultrasoft quark is more complicated, 
and it contains both a local and a nonlocal term \cite{radlept2}
\begin{eqnarray}\label{Jem}
\varepsilon^{*\mu} j_\mu^{\rm e.m.} = e_q \, 
\bar q \varepsslash^* q_{n,\omega} + 
T\{ J^{\xi\xi}\,, \int d^4 x
i{\cal L}_{q\xi}^{(1)}(x)\}
\end{eqnarray}
where $J^{\xi\xi}$ describes the coupling of a $n-$collinear photon to
$n-$collinear quarks. This scales like $O(\lambda^{-1})$ and it appears
in time-ordered products with a subleading ultrasoft-collinear Lagrangian.
It is given by
\begin{eqnarray}
J^{\xi\xi} &=& e_q \, 
\bar q_{n,\omega_1} \varepsslash^* \frac{1}{\bar n\cdot i\partial}
\big[ W^\dagger i\Dslash_{c\perp } \frac{\bnslash}{2}\xi_n \big]_{\omega_2}\\
& & +
e_q \,  
\big[\bar \xi_n i\Dslash_{c\perp } W\big]_{\omega_1} \frac{1}{\bar n\cdot i\partial}
\varepsslash^* \frac{\bnslash}{2} q_{n,\omega_2}\nn
\end{eqnarray}

Finally, we match the SCET$_{I}$ expression of the correlator~(\ref{corr})
onto SCET$_{II}$. 
After soft-collinear factorization \cite{scet3}, the
collinear quarks and gluons with virtuality $p_c^2 \sim E_\gamma
\Lambda$ decouple from the soft modes.  This leads to the final form
of the time-ordered product in Eq.~(\ref{corr}) in the effective
theory
\begin{eqnarray}\label{Jeff}
&& \varepsilon^{*\mu} T_{\mu\nu}(q) \to
- C_1^{(v)} \, e_u \, \int dk_+ J(k_+) 
O_{1\nu}(k_+)  \\
&&\! - \!
[C_2^{(v)} v_\nu + (C_1^{(v)}\! + C_3^{(v)}) \frac{n_\nu}{n\cdot v} ] e_u \!\! 
\int \! dk_+ J(k_+) O_2(k_+) \ , \nn
\end{eqnarray}
with  the two soft operators given by nonlocal quark bilinears
\begin{eqnarray}
O_{1\mu}(k_+) &=& \int \frac{d\lambda}{4\pi} e^{-\frac{i}{2} \lambda k_+} \, 
\bar q(\lambda \frac{n}{2}) Y_n(\lambda \frac{n}{2},0) 
\epsslash^*_\perp \frac{\nslash}{2}
\gamma_\mu^\perp P_L b_v(0) \nn \\
O_2(k_+) &=& \int \frac{d\lambda}{4\pi} e^{-\frac{i}{2} \lambda k_+} \, 
\bar q(\lambda \frac{n}{2}) Y_n(\lambda \frac{n}{2},0)
\epsslash^*_\perp \frac{\nslash}{2}
P_R  b_v(0) \ .  \nn \\
\label{O12}
\end{eqnarray}
Here $Y_n(x,y) = P\exp(ig\int_{y_-}^{x_-} d\alpha\, n\cdot A(\alpha n))$ is
a Wilson line along $n_\mu$.

The jet function $J$ describes the effects of the collinear quarks and
gluons with virtualities $p_c^2 \sim E_\gamma \Lambda$. It can be written as a
vacuum expectation value of SCET$_{I}$ collinear fields
\cite{radlept2}, and as such it is calculable in perturbation theory
in an expansion in $\alpha_s(\sqrt{Q\Lambda})$. The explicit result to
subleading order in
$\alpha_s$ is (with $L = \log [(-2E_\gamma k_+ - i\varepsilon)/\mu^2)$)
\begin{eqnarray}\label{Jdef}
J(k_+) = \frac{1}{k_+ + i\varepsilon} \left(1 + \frac{\alpha_s C_F}{4\pi}
(L^2 - 1 - \frac{\pi^2}{6})\right)
\end{eqnarray}
Finally, the nonperturbative physics is contained in the matrix elements
of the soft operators $O_{1\mu}(k_+)$ and $O_2(k_+)$ in Eq.~(\ref{O12}).

An alternative derivation is possible, wherein one matches directly
from QCD onto SCET$_{II}$. In this approach \cite{radlept3,qcd2scet2}, the jet 
function $J$ together with the hard coefficients $C_i^{(v)}$ are obtained
as the Wilson coefficient in the QCD $\to $ SCET$_{\rm II}$ matching. 
We will use here the 2-step approach, which has the advantage of allowing a 
transparent connection between the SCET$_I$ Wilson coefficients in 
(\ref{Jweak}) and the Wilson coefficient of the SCET$_{II}$ operator.

The discussion above parallels closely the proof of factorization in $B\to \gamma \ell\bar\nu$
decays  in Ref.~\cite{radlept2}, except that we were careful to keep it independent on the
final hadronic state, which is only assumed to be soft. 
For example, we included also the operators in the second term of Eq.~(\ref{Jeff}) 
which do not contribute to leptonic radiative B decays. 

To complete the proof of factorization one has to show that no
other operators can appear at leading order in Eq.~(\ref{Jeff}).
Since soft-collinear factorization in SCET$_{\rm I}$ preserves the form 
of the soft operators, it is clear that the T-products of the operators
contributing to the weak current 
Eq.~(\ref{Jweak}), and to the e.m. current Eq.~(\ref{Jem}), will always
reduce to combinations of $O_{1\mu}$ and $O_2$. 
Photon couplings to the spectator quarks require at least 2 insertions
of the soft-collinear Lagrangian ${\cal L}_{q\xi}^{(1)}$, so they are
power suppressed. Finally, no operators containing additional soft
quarks and gluons appear at leading order; the proof is similar to
that for $B\to \gamma e\nu$, to which such operators would also contribute.
For a recent study of these subleading contributions in the radiative
leptonic decay, see Ref.~\cite{subl}.

Taking the $B\to $ vacuum matrix element of the SCET$_{II}$ operators in 
Eq.~(\ref{Jeff}) gives directly the hadronic tensor relevant for the radiative 
leptonic $B\to \gamma \ell\nu$.  This can be evaluated explicitly
using the  $B\to 0$ matrix element of the soft operator $O_{1\mu}(k_+)$. The
latter is parameterized in terms of the B meson light-cone wave functions, 
defined as \cite{radlept3}
\begin{eqnarray}\label{wv}
&& \int \frac{d\lambda }{4\pi} e^{-\frac{i}{2} k_+ \lambda}
\langle 0 | \bar q_i(\lambda\frac{n}{2})Y_n(\lambda,0) b_v^j(0)|\bar B(v)\rangle = \\
&& -\frac{i}{4}f_B m_B \left\{
\frac{1+\vslash}{2} [\bnslash n\cdot v \phi_+(k_+) +
\nslash \bn\cdot v \phi_-(k_+) ]\gamma_5 \right\}_{ji} \nonumber
\end{eqnarray}
With this definition, the wave functions are normalized as 
$\int dk_+ \phi_\pm(k_+) = 1$.
Using the soft matrix element Eq.~(\ref{wv}) into Eq.~(\ref{Jeff})
reproduces the leading order factorization relation for the
radiative leptonic decay form factors \cite{radlept0,radlept1,radlept2,radlept3}.

Proceeding in a similar way, by taking the matrix element of the operator 
identity Eq.~(\ref{Jeff}) for the $B\to \pi$ transition (with $\pi$ a soft 
pion) gives a factorization relation for the semileptonic radiative decay
$B\to \pi\gamma e\bar\nu$. This requires the 
introduction of the soft functions $S_{1,2}(k_+, t^2, \zeta)$
defined by (with $t=p-p'$ and $\zeta = (n\cdot p')/(n\cdot v)$)
\begin{eqnarray}\label{H1def}
&& \langle \pi(p') |O_{1\mu} (k_+ ) | \bar B^0(v)\rangle \\
&& \hspace{2cm} =
- (g_{\mu\nu}^\perp + i\varepsilon_{\mu\nu}^\perp) \varepsilon^*_\nu
S_1(k_+, t^2, \zeta) \nn \\
\label{H2def}
&& \langle \pi(p') |O_2(k_+ ) |\bar B^0(v)\rangle \\
&& \hspace{2cm} =
- (g_{\mu\nu}^\perp + i\varepsilon_{\mu\nu}^\perp) \varepsilon^*_\nu p'_\mu
S_2(k_+, t^2, \zeta)\,, \nn
\end{eqnarray}
Similar nonperturbative functions appear in the theory of
off-diagonal hard-scattering exclusive processes \cite{GPD}, and are
known as GPDs (generalized parton distributions). See Refs.~\cite{GPDreviews}
for recent reviews. In the context
of $b\to u$ decays, they were considered previously in
Ref.~\cite{skewb}, and $S_1$ is directly related to the
function $\tilde {\cal F}_\zeta^{(1)}(x,q^2)$ introduced in that paper.

The soft functions $S_{1,2}$ satisfy several model-independent
constraints. T invariance and complex conjugation constrains them to 
be real. Their zeroth moments over $k_+$
are related to the usual $B\to \pi$ form factors (see the Appendix
for the definitions of the form factors)
\begin{eqnarray}
&& \int_0^\infty \mbox{d} k_+ S_1(k_+,t^2,\zeta) = 
-\frac12 n\cdot p f_+(t^2) + \frac14 f_2(t^2) n\cdot t \nn \\
&& \int_0^\infty \mbox{d} k_+ S_2(k_+,t^2,\zeta) = 
\frac14 n\cdot p f_T(t^2) \,.
\end{eqnarray}
while higher moments are related to the corresponding form factors
of higher-dimensional operators
\begin{eqnarray}
&& \int_0^\infty \mbox{d} k_+ (k_+)^N S_1(k_+,t^2, \zeta) \\
&&\qquad
= -\langle \pi(p') |\bar q (-n\cdot i {\overleftarrow D})^N \frac{\nslash}{2} P_L b_v |\bar B(p)\rangle \nn \\
&& p^{\prime \perp}_\nu
\int_0^\infty \mbox{d} k_+ (k_+)^N S_2(k_+,t^2, \zeta) \\
&&\qquad 
= - \frac12
\langle \pi(p') |\bar q (-n\cdot i {\overleftarrow D})^N \frac{\nslash}{2} \gamma_\nu^\perp b_v |\bar B(p)\rangle \nn
\end{eqnarray}

Inserting the definitions of the soft functions Eqs.~(\ref{H1def}),
(\ref{H2def}) into the operator expression for the hadronic tensor
$T_{\mu\nu}$ Eq.~(\ref{Jeff}), one finds the leading order
factorization relations listed in Eqs.~(\ref{V14})-(\ref{A14}).

Similar relations can be given for other semileptonic radiative B decays,
such as $B\to \rho \gamma e\bar\nu$ and $B\to K\gamma e^+e^-$, in terms
of appropriately defined soft functions.
In order to make detailed predictions, some information
is required about the soft functions $S_{1,2}$.
In the next section we show that in the limit of a soft pion, chiral 
symmetry fixes the functions $S_{1,2}(k_+,t,\zeta)$ 
in terms of the B meson light-cone wave function defined in Eq.~(\ref{wv}).

\subsection{Soft pion matrix elements of $O_{1\mu}, O_2$}

The $B\to \pi$ matrix elements of the soft operators 
$O_{1\mu}(k_+)$ and $O_2(k_+)$ are constrained by 
chiral symmetry and heavy quark symmetry. 
The appropriate theory tool to derive these constraints is heavy hadron chiral
perturbation theory \cite{wise,BuDo,TM,review}.  This effective theory
describes the interaction of Goldstone bosons with themselves and with 
heavy hadrons. The Goldstone bosons are contained in the nonlinear
field $\xi = e^{iM/f}$ ($\Sigma=\xi^2$), where $f$ is the pion decay
constant $f \simeq 131$~MeV, and $M$ is the usual SU(3) matrix

\begin{eqnarray}
M = \left(
\begin{array}{ccc}
 {1\over\sqrt2}\pi^0 +
{1\over\sqrt6}\eta &
\pi^+ & K^+ \\
\pi^-& -{1\over\sqrt2}\pi^0 + {1\over\sqrt6}\eta&K^0 \\
K^- &\bar K^0 &- {2\over\sqrt6}\eta \\
\end{array}
\right)
\end{eqnarray}
%These fields transform as
%\begin{eqnarray}
%\Sigma \to L \Sigma R^\dagger\,,\qquad \xi \to L \xi U^\dagger = U \xi R^\dagger
%\end{eqnarray}
%under the chiral group $SU(3)_L \times SU(3)_R$.

The velocity-dependent heavy meson fields $(B_v,B_v^*)$ are grouped into the 
superfield
\begin{eqnarray}
H^a = \frac{1+\vslash }{2} \left[ B^{*\mu}_{a} \gamma_{\mu}
- B_a \gamma_5 \right]. 
\end{eqnarray}
where $a$ is a SU(3) flavor index labelling the flavor of the spectator quark in the 
$B^{(*)}$.

\begin{figure}
\begin{tabular}{cc}
{\includegraphics[height=1.3cm]{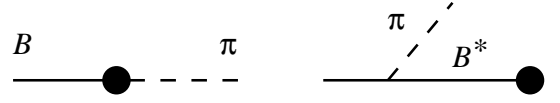}}
\end{tabular}
\caption{\label{fig2} 
Chiral perturbation theory graphs contributing to the $B\to \pi$
matrix element of the soft operators $O_{1\mu}(k_+)$ and $O_2(k_+)$
(represented by filled circles) appearing in the factorization
relation for $B\to \gamma \pi \ell\bar\nu$. The photon is not shown.
}
\end{figure}

The interactions of these fields are described by the effective Lagrangian
\cite{wise,BuDo,TM,review}
\begin{eqnarray}\label{Lag}
{\cal L} &=& \frac{f^2}{8}
\mbox{Tr } \left( \partial^{\mu} \Sigma
\partial_{\mu} \Sigma^\dagger \right) 
-i \mbox{Tr }\bar H^{a} v_{\mu} \partial^{\mu} H_a \nonumber \\
&+ &\frac{i}{2} \mbox{Tr }\bar H^{a} H_b v^{\mu} \left[ \xi^\dagger
\partial_{\mu} \xi + \xi \partial_{\mu} \xi^\dagger \right]_{ba} \nonumber\\
&+ &\frac{ig}{2} \mbox{Tr }\bar H^{a} H_b \gamma_{\nu} \gamma_5
\left[\xi^\dagger \partial^{\nu} \xi - \xi \partial^{\nu}
\xi^\dagger \right]_{ba} + \cdots 
% \nonumber 
\end{eqnarray}

The symmetries of the theory constrain also the form of operators such
as currents.  For example, the left handed current
$L^{\nu}_a = \bar q_a \gamma^\nu P_L Q$ in QCD can be written
in the low energy chiral theory as \cite{wise,BuDo,TM}
\begin{eqnarray}\label{JL}
L^{\nu}_a = \frac{i \alpha}{2}
\mbox{Tr} [\gamma^\nu P_L H_b \xi^\dagger_{ba}] + ...,
\end{eqnarray}
where the ellipsis denote higher dimension operators in the chiral and
heavy quark expansions. The parameter $\alpha$ is obtained by taking the
vacuum to B matrix element of the current, which gives
$\alpha=f_B\sqrt{m_B}$ (we use a nonrelativistic normalization for
the $|B^{(*)}\rangle$ states as in \cite{review}).

The chiral realization of the nonlocal soft operators
appearing in (\ref{Jeff}) is constructed in a similar way \cite{GPchiral}. Both 
operators $O_{1,2}(k_+)$ contain only the
left chiral light quark field, so they transform as $(\mathbf{\overline{3}}_L, \mathbf{1}_R)$
under the chiral group. Making the flavor index explicit, 
they are both of the form (see Eq.~(\ref{O12})) 
\begin{eqnarray}
&& O_\Gamma^a(k_+) = [(\bar q^a Y)_{k_+} P_R \Gamma (Y^\dagger b_v)] \ . 
\end{eqnarray}
In analogy with the local current (\ref{JL}) we write for $O_\Gamma^a(k_+)$
in the chiral theory
\begin{eqnarray}\label{JOL}
&& O^a_\Gamma(k_+)  = \frac{i}{4} 
\mbox{Tr} [\hat \alpha_L(k_+) P_R\Gamma H_b \xi^{\dagger }_{ba}],
\end{eqnarray}
where the most general form for $\hat \alpha_{L}(k_+)$ depends on four
unknown functions $a_i(k_+)$
\begin{eqnarray}
\hat \alpha_L(k_+) = a_1 + a_2 \nslash + a_3 \vslash +
\frac12 a_4 [\nslash, \vslash]
\end{eqnarray}
The heavy quark symmetry constraint $H \vslash = - H$ reduces the 
number of these functions to two. Taking the vacuum to B meson matrix element 
fixes the remaining functions as
\begin{eqnarray}\label{alpha}
\hat\alpha_L(k_+) = - f_B \sqrt{m_B} [\bnslash \phi_+(k_+) + 
\nslash \phi_-(k_+)]
\end{eqnarray} 
where $\phi_\pm(k_+)$ are the usual light-cone wave functions of a B
meson, defined in Eq.~(\ref{wv}). 

We are now in a position to compute the $B\to \pi$ matrix element of
the soft operators in Eq.~(\ref{Jeff}). There are two chiral
perturbation theory contributions shown in Fig.~2. Considering the
example of the operator $O_{1\mu}$, the two diagrams are given by
\begin{eqnarray}
&& I_a = 
%-  
\frac{1}{4f_\pi} f_B m_B \phi_+(k_+) (g^\perp_{\mu\nu} + 
i \varepsilon^\perp_{\mu\nu} ) \varepsilon^{*\nu}_\perp\\
&& I_b = 
%- 
i \frac{i}{4} f_B m_B \phi_+(k_+) 
(g^\perp_{\mu\nu} + 
i \varepsilon^\perp_{\mu\nu} ) \varepsilon^{*\nu}_\perp \\
&& \qquad \times \frac{-i(\bar n_\alpha - v_\alpha \bn\cdot v)}{2(-v\cdot p' - \Delta)}
(-) \frac{2g}{f_\pi} i p'^\alpha \nn
\end{eqnarray}
Adding them together and performing a similar calculation for
$O_2(k_+)$ gives the final result for $B\to \pi$ matrix elements
to leading order in the chiral expansion:
\begin{eqnarray}\label{O12me}
&& \langle \pi(p')|O_{1\mu}(k_+)|\bar B(p)\rangle = 
 \frac{1}{4 f_\pi} f_B m_B \phi_+(k_+) \\
&& \quad \times(g_{\mu\nu}^\perp +
i \varepsilon_{\mu\nu}^\perp) \varepsilon^{*\nu} 
\left( 1 + g \frac{e_3\cdot p'}{v\cdot p' + \Delta} \right)\nn \\
&& \langle \pi(p')|O_{2}(k_+)|\bar B(p)\rangle =
-  \frac{1}{4 f_\pi} f_B m_B \phi_+(k_+) \\
&& \quad \times p^{\prime\alpha} \varepsilon^{*\beta}  (g_{\alpha\beta}^\perp +
i \varepsilon_{\alpha\beta}^\perp) 
\frac{g}{v\cdot p' + \Delta} \nn \ , 
\end{eqnarray}
where the unit vector $e_3$ is defined as $e_3 = \frac12 (n - \bn)$.
Confronting these expressions with Eqs.~(\ref{H1def})-(\ref{H2def})
one obtains the results for $S_{1,2}$ reported in
Eqs.~(\ref{S1res})-(\ref{S2res}).

\vspace{0.5cm}

\section{Phenomenology and numerical study}
\label{sect:pheno}

In this section we present some phenomenological applications 
of our factorization formula, with a twofold intent:
(i) give predictions for quantities such as photon spectrum 
and the integrated rate with appropriate cuts;
(ii) compare our effective-theory based results with simplified approaches to 
$\bar{B} \to \pi e \bar{\nu} \gamma$, which are usually implemented 
in the Monte Carlo (MC) simulations used in the experimental analysis. 
All numerical results reported here refer to the neutral B decay
$\bar{B}^0 \to \pi^+ e^- \bar{\nu}_e \gamma$.

\subsection{Factorization results}

We present here results for the photon energy spectrum, the 
distribution in the electron-photon angle, and integrated 
rate with various kinematical cuts, forced upon us by 
the theoretical tools used:
\begin{enumerate}
\item An upper cut on the pion energy in the B rest frame 
$E_\pi < E_\pi^{\rm max}$. This restriction is required by the
applicability of the chiral perturbation theory computation of the
soft matrix elements of $O_{1\mu , 2}$.  
\item  A lower cut on the
charged lepton-photon angle in the rest frame of the $B$ meson,  
$\theta_{e \gamma} > \theta^{\rm min}_{e \gamma}$. 
This is required in  
order to regulate the collinear singularity introduced
by photon radiation off the charged lepton leg in the massless lepton limit.  
\item  Finally,  
to ensure the validity of the factorization theorem,
we require the photon to be sufficiently energetic (in the $B$ rest frame)  
$E_\gamma^{\rm min} \gg \Lambda_{QCD}$.
We choose $E_\gamma^{\rm min} = 1$ GeV.
\end{enumerate} 

In our evaluation we use the transition matrix element in 
Eq.~(\ref{full}), with the factorization expressions for the hadronic
amplitudes $T_{\mu\nu}^{V,A}$. 
Working to leading order in $\alpha_s$, the only surviving form 
factors are $V_{1,2}$ and $A_{1,2}$ and the nonperturbative input needed   
is the first inverse moment of the 
light-cone B wave functions $\phi_+^B(k_+)$:
\begin{equation}
\lambda_B^{-1}= \int dk_+  \  \frac{\phi_B^+(k_+,\mu)}{k_+} \ .
\end{equation} 
Moreover, the $\bar B^0\to \pi^+ \ell^-\bar\nu$ 
transition matrix element $F_\nu(p,p')$ is parameterized as
\begin{eqnarray}
F_\nu(p,p') = f_+(E_\pi) (p+p')_\nu + f_-(E_\pi) (p-p')_\nu
\end{eqnarray}
We adopt for this
matrix element the leading order $HH \chi PT$ result \cite{wise,BuDo,TM,review}
\begin{eqnarray}
f_+(E_\pi) = - f_-(E_\pi) = g\frac{f_B m_B}{2f_\pi} \frac{1}{E_\pi+\Delta}
\label{eq:fp}
\end{eqnarray}
The corresponding result for $B^-\to \pi^0\ell^-\bar\nu$ is multiplied by  
a factor $1/\sqrt2$.  
We summarize in Table~\ref{table} the numerical ranges used for 
the various input parameters \cite{parameters,parameters2}. 

The phase space integration with the cuts described above was
performed using the Monte Carlo event generator
RAMBO~\cite{RAMBO}. The factorization results for the photon energy
spectrum and the distribution in $\cos \theta_{e \gamma}$ are shown in
Fig.~\ref{fig:spectra2} (solid lines), using
the central and extreme values for the input parameters as reported in
Table~\ref{table}. The shaded bands reflect the uncertainty in our prediction,
which is due mostly to the variation of $\lambda_B$.
For $E_\gamma^{\rm min}=1 \ {\rm GeV}$, $E_\pi^{\rm max}=0.5 \ {\rm GeV}$, 
and $\theta_{e \gamma}^{\rm min}=5^o$, 
the integrated radiative branching fraction predicted by our factorization 
formula is (up to small perturbative corrections in $\alpha_s(m_b)$ and 
$\alpha_s(\sqrt{m_b \Lambda})$)
\begin{eqnarray}
{\rm Br}_{\bar{B} \to \pi e \bar{\nu} \gamma}^{\rm cut} ({\rm fact}) 
&=& 
 \Big(1.2 \  \pm 0.2 (g)  \ ^{+ 2.2} _{-0.6} (\lambda_B)  \Big) 
\times 10^{-6} 
\nonumber \\
&\times & \left(\frac{|V_{ub}|}{0.004}\right)^2  \times
\left(\frac{f_B}{200 \ {\rm MeV}}\right)^2  \ . 
\label{eq:brfact}
\end{eqnarray}
Apart from  the overall quadratic dependence on $V_{ub}$ and $f_B$, 
the main uncertainty of this prediction comes from the  poorly known 
nonperturbative parameter $\lambda_B$. 

%Finally, we comment on the radiative corrections to these predictions,
%which are known to one-loop order. 
The radiative corrections to these predictions are known to one-loop
order.  Their numerical impact has been computed in the radiative
leptonic decay case, including a resummation of large logarithmic
terms $\log^2(2E_\gamma/\mu)$ \cite{scet0,radlept0,radlept1,radlept3}
and $\log^2(2E_\gamma k_+/\mu^2)$ \cite{radlept3}. Their effect can be
significant, and was found to reduce the tree level result by about
30\%.  We do not include them in our results, in view of the large
uncertainty from the input parameters, and leave a study of their
effects for future work.

Finally, let us mention that radiative rate, photon spectrum and
angular distribution could be used in the future to obtain
experimental constraints on $\lambda_B$, given their strong
sensitivity to it.  This requires, however, a reliable estimate of the
impact of chiral corrections, $O(\alpha_s)$ corrections, and the
uncertainty in other input parameters.

%%%%%%%%%%%%%%%%%%%%%%%%%%%%%%%%%%%%%%%%%%%%%%%%%%%%%%%%%%%%%%%%
%A more stable quantity is the ratio of the radiative to the
%nonradiative modes (with the same cut on the pion energy) from which the
%dependence on $|V_{ub}|$ and $f_B$ cancels out. We define
%\begin{eqnarray}\label{Rdef}
%R(E_{\gamma}^{\rm min}, E_{\pi}^{\rm max}, \theta_{e\gamma}^{\rm min}) =
%\frac{\Gamma_{\rm rad}(E_\gamma > E_\gamma^{\rm min}, 
%\theta_{e\gamma}>\theta^{\rm min}_{e\gamma})}
%{\Gamma_{\rm sl}(E_\pi < E_\pi^{\rm max})}
%\end{eqnarray}
%The denominator in this relation is computed using the relation
%\begin{eqnarray}
%\frac{d\Gamma(\bar B\to \pi \ell\bar\nu)}{dE_\pi} =
%\frac{G_F^2 |V_{ub}|^2}{12\pi^3} m_B (E_\pi^2- m_\pi^2)^{\frac32} 
%|f_+(E_\pi)|^2
%\end{eqnarray}
%and the numerator is determined from the Monte Carlo computation.
%%%%%%%%%%%%%%%%%%%%%%%%%%%%%%%%%%%%%%%%%%%%%%%%%%%%%%%%%%%%%%

\begin{table}%[H] add [H] placement to break table across pages
\caption{\label{table} Input parameters used in the numerical 
computation~\cite{parameters,parameters1,parameters2}.}
\begin{ruledtabular}
\begin{tabular}{cc|cc}
$f_B$ & $(200\pm 30)$ MeV  & $f_\pi$ & $131$ MeV \\
$\lambda_B$  & $(350 \pm 150)$ MeV  &  $g$   & $0.5 \pm 0.1$  \\
$|V_{ub}|$ & $0.004$ &  &   \\
\end{tabular}
\end{ruledtabular}
\end{table}

\begin{table}%[H] add [H] placement to break table across pages
\caption{\label{table2} Branching fractions of the 
cut $\bar{B} \to \pi e \bar{\nu} \gamma$ decay 
% in units of $10^{-6}\times |V_{ub}/0.004|^2$, 
corresponding to the factorization amplitude and the IB1, IB2 cases 
defined in the text, 
for central values of the input parameters reported in Table~\ref{table}. 
%%%%%%%%%%%%%%%%%%%%%%
%The last column shows the ratio $R$
%of the radiative to nonradiative branching fractions (defined in 
%Eq.~(\ref{Rdef})).
%%%%%%%%%%%%%%%%%%%%%%%
}
\begin{ruledtabular}
\begin{tabular}{c|ccc}
% &  &  &  \\
%\hline
%cuts  & Br(IB1)  & Br(FACT) & $R$  \\
cuts  & $10^6$  Br(fact)  & $10^6$ Br(IB1) & $10^6$ Br(IB2)  \\
\hline
$E_\gamma > 1$ GeV &       &        &  \\
$E_\pi < 0.5$ GeV & $1.2$ & $2.4$ & $2.8$ \\
$\theta_{e\gamma}>5^\circ$ &       &        &  \\
\end{tabular}
\end{ruledtabular}
\end{table}

\subsection{Comparison with simplified approaches}

\begin{figure*}[!t]
\centering
\begin{picture}(300,170)  
\put(-10,55){\makebox(50,50){\epsfig{figure=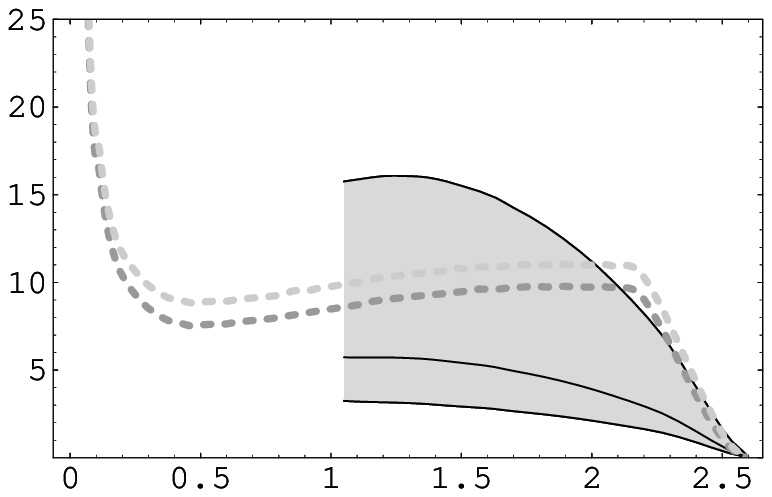,height=5cm}}}
\put(263,52){\makebox(50,50){\epsfig{figure=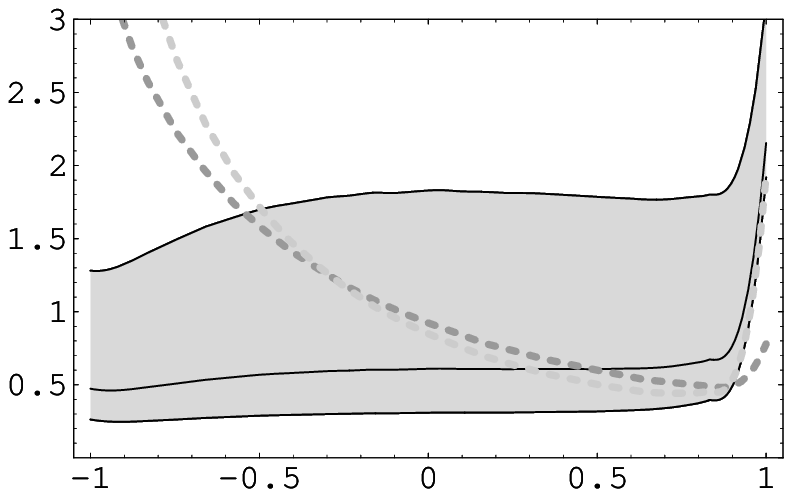,height=5cm}}}
\put(-40,155){
\framebox[1.0\width][c]{
\small $ E_\pi < 0.5 \ {\rm GeV} \qquad  \theta_{e \gamma} > 5^o  $  
}}
\put(230,155){
\framebox[1.0\width][c]{
\small $ E_\pi < 0.5 \ {\rm GeV} \qquad  E_\gamma > 1 \ {\rm GeV}  $  
}}
\put(-115,85){
{\Large 
$\frac{d \bar{\Gamma}}{d E_\gamma}$
}
}
\put(128,85){
{\Large 
$\frac{d \bar{\Gamma}}{d (\cos \theta_{e \gamma})}$
}
}
\put(280,-5){
{\large 
$\cos \theta_{e \gamma}$ 
}
}
\put(-5,-5){
{\large 
$E_\gamma$ (GeV) 
}
}
\put(-20,72){
{\small
IB2
}
}
\put(-35,48){
{\small
IB1
}
}
\put(30,50){
{\small
FACT  
}
}
%%%%%%%%%%%%%%%%%%%%%%%%%%%%%%%%%%%%%%%%%%%%%%%%%%%%%%%%%%%%%
\put(220,45){
{\small
FACT
}
}
\put(210,90){
{\small
IB1
}
}
\put(245,100){
{\small
IB2
}
}
\end{picture}
\caption{\label{fig:spectra2} 
Left panel: Differential rate 
$d \bar{\Gamma}/d E_\gamma$, as a function of the photon energy in the
$B$-meson rest frame.  The normalization is given by $\bar{\Gamma}=
10^6 \, \Gamma_{B^0}^{-1} \times \Gamma (\bar{B} \to \pi e \bar{\nu}
\gamma)$.  Right panel: Differential radiative rate 
$d \bar{\Gamma}/d ( \cos \theta_{e \gamma} )$, as a function of the
cosine of electron-photon angle in the  $B$-meson rest frame.
The shaded bands and the superimposed continuous lines
represent the uncertainty in the factorization
result (FACT) induced by the parameter $\lambda_B$, assumed to range
between 200 (upper line) and 500 (lower line) MeV, 
with all other parameters equal to the central values of Table~\ref{table}. 
The dark(light)-grey dashed line corresponds
to the IB1 (IB2) results (see text for their definition).  
}
\end{figure*}

In the Monte Carlo simulations used in most experimental analyses of B
decays, simplified versions of the radiative hadronic matrix elements
are implemented. These rely on universal properties of radiative
amplitudes following from gauge invariance, which allow one to derive in a
model-independent way the terms of order $q^{-1}$ and $q^0$ of the
radiative amplitude (see Appendix~\ref{sect:appLow} for a summary of
these results, based on Low's theorem~\cite{Low}).
These simplified amplitudes do not take into account hadronic strucure
effects and their validity is restricted in general to
emission of non-energetic photons (whose wavelength does not resolve
the structure of hadrons involved in the process, 
$E_\gamma \ll \Lambda$).
They are nonetheless used over the entire kinematical range, for 
lack of better results. 
Our interest here is to evaluate how these methods fare in the region
of soft pion and hard photon (certainly beyond their regime of
validity), where we can use the QCD-based factorization results as 
benchmark.  

We use two simplified versions of the $\bar{B} \to \pi e \bar{\nu}
\gamma$ matrix element, which we denote  IB1 and IB2 (Internal
Bremsstrahlung 1 and 2). They arise as special cases of  Low's amplitude 
that we report in Appendix~\ref{sect:appLow}:  
\begin{itemize}
\item[IB1:] This is the $O(q^{-1})$ component of Low's amplitude
(first two term in the first line of Eq.~(\ref{eq:low1})), given by
the product of the non-radiative amplitude with an appropriate
radiative term. 
The PHOTOS Monte Carlo~\cite{photos} implements this form of the
radiative amplitude.
\item[IB2:] This corresponds to using Low's amplitude of
Eq.~(\ref{eq:low1}) and setting to zero the derivatives of non-radiative
form factors (i.e. just keep the first line in Eq.~(\ref{eq:low1})).  This
type of amplitude has been used in a series of papers on $K_{\ell 3
\gamma}$ by Ginsberg~\cite{Ginsberg}.  It is often used in B-physics
applications by simply rescaling the appropriate meson masses. 
\end{itemize}
In both cases the only input needed to perform the computation is
$f_+(E_\pi)$ (see Eq.~(\ref{eq:fp})). 

The results for the photon energy spectrum and the distribution in
$\cos \theta_{e \gamma}$ are shown in Figs.~\ref{fig:spectra2} 
(IB1: dark-grey dashed lines, IB2: light-grey
dashed lines).  The integrated branching fractions are listed in
Table~\ref{table2} (together with the factorization prediction) for
central values of the input parameters given in Table~\ref{table}.
The IB1 and IB2 branching fractions scale very simply with the input
parameters, being proportional to $(|V_{ub}| \, g \, f _B)^2$.

As can be seen from the comparison of differential distributions and
BRs, the two cases IB1 and IB2 lead to almost identical predictions,
not surprisingly (and in agreement with previous claims that
PHOTOS-type distributions are in agreements with Ginsberg's ones
~\cite{RadCorr,photos}).
Both IB1 and IB2 amplitudes, however, produce results quite different
from the factorization ones, as can be seen from
Fig.~\ref{fig:spectra2} and Table~\ref{table2}.  For central values of
the input parameters reported in Table~\ref{table} the IB1 and IB2
predict a radiative BR roughly a factor of two 
larger than the factorization
result. Moreover, the shape of the distributions in $E_\gamma$ and
$\cos \theta_{e \gamma}$ are quite different, 
as can be seen from Fig.~\ref{fig:spectra2}.

There is only one corner of parameter space where the integrated rate
from factorization gets close to the IB1-IB2 result, namely for
$\lambda_B \sim 200$ MeV.  Even in this case, however, the
distributions in photon energy and electron-photon angle clearly
distinguish the two scenarios.  This is explicitly shown in
Fig.~\ref{fig:spectra2}, where the uncertainty in the factorization
result induced by $\lambda_B$ is explicitly displayed through the shaded
region.

We conclude that, in the kinematical region of soft pions and hard
photons, the QCD-based factorization predictions for the radiative
decay are quite different from the ones based on the simplified
amplitudes IB1 and IB2 (PHOTOS-type and Ginsberg-type). 
We hope that our results can help construct more reliable MCs for
the simulation of $\bar{B} \to \pi e \bar{\nu}\gamma$ decays, 
or at least address the systematic uncertainty to be assigned to the 
present MC results.  
The codes used to generate the distributions given in this section are
available from the authors upon request.

\section{Conclusions}
\label{sect:conclusions} 

We derived in this paper a new factorization relation for the
exclusive semileptonic radiative $B\to \gamma \pi \ell\nu$ decays, valid in the
kinematical region with a soft pion and an energetic photon.
In this region, the form factors satisfy the factorization relation written 
in schematic form in Eq.~(\ref{1}). 
The factorization relation is easily generalized to any process 
$B\to \gamma M\ell \nu$, with $M = \rho, \omega, \cdots$ a light soft meson.

In all cases, this relation contains as input parameters the
nonperturbative matrix elements of two soft operators for the $B\to M$ 
transition. These matrix elements are the B physics analog of the  
generalized parton distribution functions encountered in the  
factorization relation for the off-diagonal $\gamma^* N\to N'\gamma$
transition, or 'deeply virtual Compton scattering' (DVCS)
\cite{GPDreviews}.

For sufficiently slow pions, the $B\to \pi$ soft function can be computed
using chiral perturbation theory methods. Previous applications of chiral
symmetry to constrain GPDs for light quark systems have been given in 
\cite{DVCSchiral}.
At leading order in $\Lambda/m_b$ 
and in the chiral expansion, the only nonperturbative quantity required is
the B meson light-cone wave function.

We have presented phenomenological applications of our factorization
formula, with two main goals in mind.  On one hand, we have given
factorization predictions for radiative observables (such as photon
spectrum and integrated rate) that can be tested as increasing
statistics becomes available at the B factories.  Such a test would
provide an important validation of the theoretical tools used 
here.  On the other hand, we have compared our results with simplified
approaches to radiative hadronic decays, relying on universal
properties of radiative amplitudes following from gauge invariance,
not taking into account hadronic strucure effects.  These simplified
amplitudes are currently implemented in the MC simulations used in
experimental analysis.  In the kinematical region accessible to our
theoretical tools (energetic photon and soft pion in the $B$ rest
frame), we find a large discrepancy between our factorization results
and PHOTOS-type or Ginsberg-type results.  
We hope our work will provide a theoretical basis to construct
improved MCs for the simulation of $\bar{B} \to \pi e
\bar{\nu}\gamma$, which is a crucial ingredient in 
a high precision measurement of the semileptonic mode $\bar{B}\to \pi e
\bar{\nu}$.

\begin{acknowledgments}
We thank Iain Stewart and Chris Lee for useful discussions and
comments on the manuscript.  We thank the Institute for Nuclear Theory
at the University of Washington for its hospitality and the Department
of Energy for partial support during the completion of this work.
This work was supported by Caltech through the Sherman Fairchild fund
(VC), and by the DOE under the cooperative research agreement
DOE-FC02-94ER40818 (DP).
\end{acknowledgments}
%\vspace{0.5cm}

\appendix

\section{Low's amplitude}
\label{sect:appLow}

In this Appendix we report the result for $A ( \bar{B}(p) \to \pi(p')
 e^-(p_e) \bar{\nu}(p_\nu) \gamma(q,\varepsilon) ) $ predicted at small
 $q$ by Low's theorem~\cite{Low}. This serves as basis of simplified
 treatments of the radiative process~\cite{photos,Ginsberg}, and is
used in the text for a comparison with the results of our paper following
from the factorization relation.

Assuming $E_\gamma \ll \Delta = m_{B^*}-m_B$, one finds for the neutral 
mode decay: 
\begin{widetext}
\begin{eqnarray}
\label{eq:low1}
A^{\bar{B}^0 \pi^+}_{\rm Low} & = & 
\frac{e G_F}{\sqrt{2}} V_{u b}  \, 
\bar{u} (p_e) \Big[ 
\Big( - \frac{\varepsilon^* \cdot p'}{q \cdot p'} 
+ \frac{\varepsilon^* \cdot p_e}{q \cdot p_e}  + 
\frac{\varepsslash^* \qslash}{2 q \cdot p_e}  
\Big) 
T^0(p,p') 
\Big]
v(p_\nu)    
\nonumber \\
&+&  \frac{e G_F}{\sqrt{2}} V_{u b}  \, 
\bar{u} (p_e) \Big[ 
2 p \cdot D(p') \Big( 
2 \pslash  \frac{\partial f_+}{\partial t^2} - 
m_e \frac{\partial f_2}{\partial t^2} \Big)  (1 - \gamma_5) 
\Big]
v(p_\nu)    \ , 
\end{eqnarray}
where the form factors $f_\pm(t^2)$ appear in the parameterization of
the matrix element $F_\mu (p,p')$ given in Eq.~(\ref{FFdef}) and are
defined as
\begin{eqnarray}
F_{\mu} (p,p') = f_+(t^2) (p + p')_\mu + f_-(t^2) (p - p')_\mu 
\end{eqnarray}
with $t = p - p'$. 
We define also
$f_{1,2} (t^2) = f_+ (t^2) \pm f_-(t^2)$, 
and 
\begin{eqnarray}
T^0 (p,p') &=& [ 2 \pslash f_+ (t^2) - m_e  f_2 (t^2) ]  (1 - \gamma_5) \ , \\
D (v)%_\mu
&=&  \frac{\varepsilon^* \cdot v}{q \cdot v} q%_\mu 
- 
\varepsilon^*%_\mu 
\ . 
\end{eqnarray}
\end{widetext}
For reference, we quote here also the definition of the 
$B\to \pi$ tensor form factor, used in Sec.~IV
\begin{eqnarray}
\langle \pi(p') |\bar q i\sigma_{\mu\nu} b |\bar B(p)\rangle =
f_T(t^2) (p_\mu p^\prime_\nu - p_\nu p^\prime_\mu )
\end{eqnarray}

In the limit $m_e \to 0$ Low's amplitude leads to the following
expressions for the invariant form factors appearing in
Eq.~(\ref{full}):
\begin{widetext}
\begin{eqnarray}
V_1 &=& f_2(t^2) \nonumber \\
V_{2,3,4} &=&  - \frac{2 f_+(t^2)}{q \cdot p'} \, \left[ 
1 - 2 \, q \cdot W  \frac{1}{f_+(t^2)} \frac{\partial f_+}{\partial t^2}
\right] \nonumber \\
A_{1,2,3,4}&=&0 
\end{eqnarray}

Similarly, for the charged mode decay we find:
%\begin{widetext}
\begin{eqnarray}
A^{B^- \pi^0}_{\rm Low} & = & 
\frac{e G_F}{\sqrt{2}} V_{u b}  \, 
\bar{u} (p_e) \Big[ 
\Big( - \frac{\varepsilon^* \cdot p}{q \cdot p} 
+ \frac{\varepsilon^* \cdot p_e}{q \cdot p_e}  + 
\frac{\varepsslash^* \qslash}{2 q \cdot p_e}  
\Big) 
T^-(p,p') 
\Big]
v(p_\nu)    
\nonumber \\
&+&  \frac{e G_F}{\sqrt{2}} V_{u b}  \, 
\bar{u} (p_e) \Big[ 
2 p' \cdot D(p) \Big( 
2 \pslash'  \frac{\partial f_+}{\partial t^2} + 
m_e \frac{\partial f_1}{\partial t^2} \Big)  (1 - \gamma_5) 
\Big]
v(p_\nu)    \ , 
\end{eqnarray}
\end{widetext}
where
\begin{equation}
T^- (p,p') = [ 2 \pslash' f_+ (t^2) + m_e  f_1 (t^2) ]  (1 - \gamma_5) \ . 
\end{equation}
\vspace{1.0cm}

\end{document}